\documentclass[12pt]{article}
\hoffset= - 2cm
\usepackage{amssymb,graphicx}
\usepackage{epstopdf}
\usepackage{amsmath,amsfonts}
\usepackage[normalem]{ulem}
\usepackage{epsfig}
\usepackage[dvipsnames]{xcolor}

\newcommand{\be}{\begin{equation}}
\newcommand{\ee}{\end{equation}}
\newcommand{\bea}{\begin{eqnarray}}
\newcommand{\eea}{\end{eqnarray}}

\def\tz{\tilde{\zeta}}

\usepackage{hyperref}
\definecolor{linkcolor}{HTML}{799B03}
\definecolor{urlcolor}{HTML}{799B03}
\hypersetup{pdfstartview=FitH,linkcolor=linkcolor,urlcolor=urlcolor,colorlinks=true}

\makeatletter
\newcommand*{\myfnsymbolsingle}[1]{%
  \ensuremath{%
    \ifcase#1
    \or 
      *%
    \or 
      \dagger
    \or 
      \ddagger
    \or 
      1
    \or 
      2
    \or
      3
    \or
      4
    \or
      5
    \or
      6
    \or
      7
    \or
      8
    \else 
      \@ctrerr
    \fi
  }%
}
\makeatother

\usepackage{alphalph}
\newalphalph{\myfnsymbolmult}[mult]{\myfnsymbolsingle}{}
\renewcommand*{\thefootnote}{%
  \myfnsymbolmult{\value{footnote}}%
}

\begin{document}

\begin{flushright}
INR-TH-2020-042
\end{flushright}

\begin{center}
  {\LARGE \bf  Superluminality in DHOST theory\\
    with extra scalar}

\vspace{10pt}

\vspace{20pt}
S. Mironov$^{a,c,d,e}$\footnote{sa.mironov\_1@physics.msu.ru},
V. Rubakov$^{a,b}$\footnote{rubakov@inr.ac.ru},
V. Volkova$^{a}$\footnote{volkova.viktoriya@physics.msu.ru}
\renewcommand*{\thefootnote}{\arabic{footnote}}
\vspace{15pt}

$^a$\textit{Institute for Nuclear Research of the Russian Academy of Sciences,\\
60th October Anniversary Prospect, 7a, 117312 Moscow, Russia}\\
\vspace{5pt}

$^b$\textit{Department of Particle Physics and Cosmology, Physics Faculty,\\
M.V. Lomonosov Moscow State University,\\
Vorobjevy Gory, 119991 Moscow, Russia}

$^c$\textit{Institute for Theoretical and Experimental Physics,\\
  Bolshaya Cheryomushkinskaya, 25, 117218 Moscow, Russia}

$^d$\textit{Moscow Institute of Physics and Technology,\\
Institutski pereulok, 9, 141701, Dolgoprudny, Russia}

$^e$\textit{Institute for Theoretical and Mathematical Physics,\\
M.V. Lomonosov Moscow State University, 119991 Moscow, Russia}
\end{center}

\vspace{5pt}

\begin{abstract} 
We consider 
DHOST Ia theory interacting gravitationally with an 
additional conventional scalar field 
minimally coupled to gravity.
At the linearized level of perturbations
about cosmological background,
we find that in the presence of {a}
slowly rolling extra scalar field,
one of the modes
generically propagates at superluminal speed. This result is valid 
for any stable cosmological background.
We identify a subclass of DHOST~Ia theories in which this
superluminality property is absent, and all modes may propagate
(sub)luminally.
We discuss possible implications for the interacting 
DHOST Ia theories.

\end{abstract}

\section{Introduction}

Scalar-tensor theories have become one of the most popular 
frameworks for addressing  cosmological challenges, 
from the late-time accelerated expansion to 
the early-time dynamics of the Universe. 
The most general known type of the scalar-tensor theories, 
featuring the {desired} $2+1$ degrees of freedom,
is the Degenerate Higher-Order 
Scalar-Tensor (DHOST) theories~\cite{DHOST1,DHOST1_EST,DHOST2,DHOST3} 
(see also reviews~\cite{DHOSTReview,KobaReview}). 
DHOST theories involve second derivative terms in the Lagrangian 
and, thus,
are {superficially} described by a set of fourth
order differential 
equations of motion. However, DHOST theories are
protected against 
the Ostrogradsky ghost 
by a degeneracy property, 
which ensures that
the set of equations of motion can be combined into
a system of second order equations. 

A complete classification of DHOST theories is
given in Refs.~\cite{DHOST1_EST,DHOST2,DHOST3}. 
The most promising from the phenomenological point of view is the
so-called ''Ia'' (or ''N-1'') class~\cite{DHOST2}.
Interestingly, DHOST~Ia family
includes both Horndeski~\cite{Horndeski,GeneralizedGalileon}
and beyond Horndeski theories 
(or GLPV)~\cite{GLPV,Gleyzes} 
as its special cases.
General  DHOST Ia family and its 
Horndeski and beyond Horndeski subclasses  provide a rich framework
for cosmological model building, especially because
they shed new light on various stability
problems~\cite{LMR,Koba_nogo,Creminelli,RomaBounce,DHOST_EFT,DHOST_cosmo};
they also appear promising from the viewpoint of constructing space-times with {traversable}
wormholes~\cite{Mironov:2018pjk,Franciolini:2018aad}.

Apart from the stability issues,  modified 
gravity and, in particular, DHOST theories may suffer 
from superluminality~\footnote{To avoid confusion,
  we note that superluminality we consider in this paper is not
  parametrically suppressed and persists
  at high spatial momenta, unlike superluminality in conventional
  EFT of gravity~\cite{deRham:2019ctd} where it is harmless from the causality
  viewpoint~\cite{deRham:2020zyh}.}.
This problem has been addressed from various perspectives, e.g.,  
in Refs.~\cite{BabVikMukh,gen_original,subl_gen,MatMat,Unbraiding}.
Superluminal propagation was argued to be 
troublesome, 
since it indicates that a Lorentz-covariant UV-completion 
is impossible for such a theory~\cite{superlum1}.
The potential superluminality problem becomes indeed acute  in DHOST theories
as 
soon as one adds an extra matter component.
This was first illustrated in Ref.~\cite{MatMat}
within the cosmological Genesis scenario 
based on a specific cubic Horndeski model: 
the {scalar} perturbation
necessarily became superluminal 
in some region of  phase space
upon adding  perfect fluid 
into a formerly non-superluminal (and stable) setup.

Another result in this direction has been recently obtained in
Ref.~\cite{Mironov_subl_fluid}, 
where beyond Horndeski theories
were analysed in cosmological backgrounds
from the superluminality viewpoint.
It has been shown that
adding even the tiniest amount of  perfect
fluid with the flat-space sound speed equal to that of light,
 $c_m =1$,  
inevitably results in
the appearance of a superluminal mode (this does not necessarily
happen
for $c_m$ substantially smaller than 1). This 
applies to any stable cosmological 
background and any beyond Horndeski Lagrangian. This finding has been
supported by similar result~\cite{Mironov_subl_kess}  in the case where
{instead of perfect fluid,}
a conventional, minimally coupled scalar field 
(whose flat-space propagation is luminal, $c_m=1$) is
 added 
to a cosmological setup in beyond Horndeski 
theory.

The purpose of this paper is to show that
superluminality is quite general propery of  DHOST~Ia
theories gravitationally interacting with conventional scalar 
field(s) minimally coupled to gravity.
Namely, we prove that unless an additional constraint is
imposed on the functions in the DHOST{~Ia} Lagrangian, there exists a
superluminal mode in the scalar sector
in an arbitrary stable cosmological background with small but non-zero
kinetic energy density of the extra scalar.
{We identify an exceptional subclass of
  DHOST~Ia theories where this property does not
hold. It} is defined by the constraint~\eqref{nov18-20-5} below;
there, all modes may be safely (sub)luminal. Clearly, this subclass is
particularly interesting from the viewpoint of cosmological
model building.


\section{DHOST~Ia with extra scalar}
\label{sec:setup}

We focus on the quadratic DHOST theories
whose Jordan frame action has the following general form:
\be
\label{eq:DHOST}
\mathcal{S}_{\pi} = \int d^4x \sqrt{-g}
\left( F(\pi,X) + K(\pi,X)\Box\pi + F_2(\pi,X) R + 
\sum_{i=1}^5 A_i(\pi,X) L_i \right),
\ee
with
\be
L_1 = \pi_{;\mu\nu} \pi^{;\mu\nu}, \quad
L_2= \left(\Box{\pi}\right)^2, \quad
L_3= \pi^{,\mu} \pi_{;\mu\nu} \pi^{,\nu} \Box{\pi}, \quad
L_4= \pi^{,\mu} \pi_{;\mu\nu} \pi^{;\nu\rho} \pi_{,\rho}, \quad
L_5= \left( \pi^{,\mu} \pi_{;\mu\nu} \pi^{,\nu} \right)^2,
\ee
where $\pi$ is a scalar field,
$X=g^{\mu\nu}\pi_{,\mu}\pi_{,\nu}$,
$\pi_{,\mu}=\partial_\mu\pi$,
$\pi_{;\mu\nu}=\nabla_\nu\nabla_\mu\pi$,
$\Box\pi = g^{\mu\nu}\nabla_\nu\nabla_\mu\pi$.
In what follows we restrict our analysis to the DHOST Ia class, which
is the most phenomenologically viable~\cite{DHOST_EFT}. 
In this class, the functions $A_2$, $A_4$ and $A_5$ 
are expressed through the independent functions
$F_2$, $A_1$ and $A_3$ as follows~\cite{DHOSTReview}:
\begin{subequations}
  \label{nov29-20-1}
  \begin{align}
A_2 &= - A_1\\
\label{a4_A}
A_4&= \frac{1}{8(F_2-X A_1)^2}\left[-16 X A_1^3+4 (3F_2+16 XF_{2X})A_1^2
-X^2 F_2 A_3^2\qquad 
\right.
\cr
&\qquad 
\left.
-(16X^2 F_{2X}-12XF_2) A_3 A_1
-16 F_{2X}(3F_2+4XF_{2X})A_1
\right.
\cr
&\qquad 
\left.
+8F_2 (XF_{2X}-F_2)A_3+48F_2 F_{2X}^2\right]
\\
\label{a5_A}
A_5&=\frac{\left(4F_{2X}-2A_1+XA_3\right)\left(-2A_1^2-3XA_1A_3+4F_{2X} A_1+4F_2 A_3\right)}{8(F_2-XA_1)^2}\,.
  \end{align}
  \end{subequations}
Other arbitrary functions $F$ and $K$ belong to  Horndeski subclass
and do not enter these relations.

Our starting point is (perturbed)
spatially flat FLRW background with
metric 
{ \be
  \mathrm{d}s^2
= (1+\alpha)^2 \mathrm{d}t^2 -
\gamma_{ij}(\mathrm{d}x^i+ \partial^i\beta \mathrm{d}t)(\mathrm{d}x^j
+\partial^j\beta \mathrm{d}t) \; ,
\label{nov26-20-1}
\ee
with
\be
\gamma_{ij}= a^2(t) e^{2\zeta} \left(\delta_{ij} + h_{ij}^T +
\dfrac12 h_{ik}^T {h^{k\:T}_j}\right)\; ,
\label{nov26-20-2}
\ee
where 
$\alpha$, $\beta$ and $\zeta$ are lapse, shift and curvature perturbations, respectively, 
and $h_{ij}^T$ is traceless and
transverse perturbation.}
It is
supported by {a} rolling DHOST field $\pi=\pi (t)$.
We assume that this background is stable, and DHOST perturbations about
it are not superluminal. We neither  impose any
further constraints on the background 
nor {assume any relations
  other than \eqref{nov29-20-1} between}
  the Lagrangian functions in~\eqref{eq:DHOST} yet.
In the unitary gauge
$\delta \pi =0$, 
{dynamical}
perturbations 
in the DHOST sector divide into tensor modes
$h_{ij}^T$ and the scalar mode 
$\tilde{\zeta}$ (explicitly defined below).
In pure DHOST~Ia theory, the unconstrained second
order action reads 
\begin{equation}
\label{eq:quadr_action_DHOSTIa-noscalar}
\begin{aligned}
  S^{(2)}_{\pi}=
  \int\mathrm{d}t\,\mathrm{d}^3x \,a^3
  \Bigg[\left(\dfrac{\mathcal{{G}_T}}{8}\left(\dot{h}^T_{ik}\right)^2-
    \dfrac{\mathcal{F_T}}{8a^2}\left(\partial_i h_{kl}^T\right)^2\right) +
 \left( \mathcal{{G}_S} \dot{\tilde{\zeta}}^2
- \dfrac{1}{a^2}  \mathcal{{F}_S} (\partial_i \tilde{\zeta})^2 \right)
\Bigg] \; ,
\end{aligned}
\end{equation}
where
\begin{subequations}
\label{eq:App_list_of_coefficients_1}
\begin{align}
 \mathcal{G_T} &= -2 F_2 + 2 A_1 X, \label{eq:App_Gt}\\ 
 \mathcal{F_T} &= -2 F_2,\label{eq:App_Ft}
  \\
  \mathcal{G_S} & = \dfrac{\tilde{\Sigma}\mathcal{{G}_T}^2}{\tilde{\Theta}^2}
    +3\mathcal{{G}_T}, \label{eq:App_Gs}
  \\
  \mathcal{F_S} &= \dfrac{1}{a}\dfrac{\mathrm{d}}{\mathrm{d}t}
  \left[ \dfrac{a \;\mathcal{{G}_T}\left(\mathcal{G_T} + \mathcal{D} \dot{\pi} + \mathcal{F_T}\Delta\right)}{\tilde{\Theta}}\right]
  -\mathcal{F_T},
   \label{eq:App_Fs}
\end{align}
\end{subequations}
with {
\begin{subequations}
  \begin{align}
    & \mathcal{D} = -2 A_1 \dot {\pi} + 4 F_ {2 X} \dot {\pi},
    \label{calD-def}\\
&\Delta =  \frac{X}{2 \mathcal{G_T}}\left(
    2 A_1- 4 F_{2X} - A_3 X \right) \; .
    \label{nov19-20-1}
  \end{align}
\end{subequations}
Expressions for $\tilde{\Sigma}$ and $\tilde{\Theta}$ are
cumbersome and not illuminating;
they are} given in Appendix, where we derive the above formulas.
In these notations the scalar perturbation is
$\tilde{\zeta} = \zeta - \alpha \Delta$.
We note in passing that
hereafter
we do not use the background equations of motion when deriving
the action for perturbations.

 The stability of the background
requires  
$\mathcal{G_T},\mathcal{F_T},\mathcal{G_S},\mathcal{F_S}
  > 0$,
   while the sound speeds squared are
  \be
  \label{nov9-20-2}
  c_{\mathcal{S} , 0}^2 =  \frac{\mathcal{F_S}}{\mathcal{{G_S}}} \; , \;\;\;\;
  c_{\mathcal{T}}^2 =  \frac{\mathcal{F_T}}{\mathcal{{G_T}}}\;.
  \ee
  We assume that both speeds are not superluminal,
  $ {c_{\mathcal{S},0}}, c_{\mathcal{T}}
  \leq 1$.

We  add to the theory 
another, conventional 
scalar 
field $\chi$, which does not interact directly with the DHOST scalar $\pi$ and
{has minimal} coupling to gravity:
$\mathcal{S}= \mathcal{S}_{\pi} + \mathcal{S}_{\chi}$, where
\be
\label{nov7-20-1}
\mathcal{S}_{\chi} =
\int\mathrm{d}^4x\sqrt{-g} \left( \frac{1}{2} g^{\mu \nu} \chi_{,\mu}
\chi_{,\nu} - V(\chi) \right) \; .
\ee
In what follows, it is instructive to consider a somewhat more general case
in which the additional scalar field is of k-essence type~\cite{kess}:
\be
\label{eq:action_setup_kess}
S_{\chi} = \int\mathrm{d}^4x\sqrt{-g} \,
P(\chi,Y), \quad Y = g^{\mu\nu}\chi_{,\mu}\chi_{,\nu}\,.
\ee  
In the absence of DHOST field, the stability conditions for k-essence
in
FLRW background with 
$Y=\dot{\chi}^2 \neq 0$ 
have the standard form
$P_Y > 0$, $Q \equiv P_Y +2 Y P_{YY} >0$,
while the propagation speed squared of perturbations
is
\be
c_m^2 = \frac{P_Y}{Q}\;.
\label{nov9-20-3}
\ee
In the case of {a} conventional scalar $\chi$
described by the action \eqref{nov7-20-1},
which is of primary
interest, one has $c_m=1$.

\section{Rolling scalar and superluminality}

We now consider 
DHOST~Ia plus extra scalar theory
and
the background in which both $\dot{\pi}$ and  $\dot{\chi}$ do not vanish
(in particular, $Y= \dot{\chi}^2\neq 0$). The expressions for
$\mathcal{{G}_T}$ and $\mathcal{F_T}$ do not get modified,
so the
tensor
perturbations do not become superluminal. On the contrary,
the situation in the scalar sector changes dramatically.
The non-vanishing background $\dot{\chi}$ induces mixing between
the scalars $\tilde{\zeta}$ and $\delta \chi$~\cite{DHOST_EFT}, 
so the {unconstrained} quadratic
action {in the scalar sector}
reads (modulo terms with less than two derivatives,
{see
the complete expression in Appendix})
\begin{equation}
\label{eq:quadr_action_DHOSTIa+k}
\begin{aligned}
  S^{(2) \, scalar}_{\pi+\chi}
  =\int\mathrm{d}t\,\mathrm{d}^3x \,a^3\Bigg[
    G_{AB} \, \dot{v}^A \dot{v}^B 
- \dfrac{1}{a^2} F_{AB} \, \partial_i\,{v^A} \partial_i\,{v^B}
\Bigg]
,
\end{aligned}
\end{equation}
where $A,B=1,2$, 
$v^1 = \tz$, $v^2 = \delta\chi$. 
The matrices $G_{AB}$ and $F_{AB}$
have the following forms:
\be
\label{nov9-20-1}
G_{AB} = 
\begin{pmatrix}
\mathcal{G_S} + \dfrac{\mathcal{{G}_T}^2}{\tilde{\Theta}^2} YQ 
& \dot{\chi} Q g
 \\ \\
 \dot{\chi} Q g
 & Q
\end{pmatrix}, ~~~~~~~~~
F_{AB} = 
\begin{pmatrix}
\mathcal{F_S} 
&  \dot{\chi} P_Y f
\\ \\
 \dot{\chi} P_Y f & 
 P_Y
\end{pmatrix}\; ,
\ee
where
\begin{subequations}
  \begin{align}
    g &= -\dfrac{\mathcal{{G}_T}}{\tilde{\Theta}}  \left(1-3\dfrac{P_Y}{Q}\Delta
    \right),
    \\
    f &=  -\dfrac{\left(\mathcal{G_T} + \mathcal{D} \dot{\pi} + \mathcal{F_T}
      \Delta
      \right)}{\tilde{\Theta}} \; .
  \end{align}
  \end{subequations}
These expressions are valid for any $Y$.
The two sound speeds squared are eigenvalues of the
matrix $G^{-1} F$, i.e., they obey
\be
\label{eq:det}
\mbox{det}\left(F_{AB} - c_{\mathcal{S}}^2 G_{AB}\right) = 0 \; .
\ee
We begin with the general case with $f\neq g$ and take
$Y$ to be small. We have to distinguish  two situations.
(i) If $c_m^2 \neq c_{\mathcal{S} , 0}^2$
(i.e., $\ P_Y/Q
\neq  \mathcal{F_S}/\mathcal{G_S}$),
then one of the sound speeds is $ c_{\mathcal{S} , -} =
c_{\mathcal{S} , 0} + \mathcal{O}(Y)$, while the other is given by
  \be
  c_{\mathcal{S} , +}^2 = c_m^2
  \left(1 + \frac{Y(f-g)^2}{\mathcal{G_S}(c_m^2 - c_{\mathcal{S} , 0}^2)}
  \right)  + \mathcal{O}(Y^2)
  \; .
  \label{nov18-20-1}
  \ee
  This means, in particular, that in the theory 
  of conventional scalar field with $c_m=1$ and subluminal
  DHOST~{Ia}
  with  $c_{\mathcal{S} , 0} < 1$, the mode which is predominantly
  $\delta \chi$ becomes superluminal at small but non-zero
  background values of $Y$. (ii) For
  $c_m^2 = c_{\mathcal{S} , 0}^2 $, the sound speeds are given by
  \be
  c_{\mathcal{S} , \pm}^2  = c_m^2 \left[1 \pm
  \left(\frac{Y P_Y(f-g)^2}{\mathcal{G_S}} \right)^{1/2}\right]
+ \mathcal{O}(Y)
\; ,
\label{nov18-20-2}
  \ee
  which again shows that  in the case
  of conventional scalar field with $c_m=1$ one 
  of the modes becomes superluminal at small $Y$.

  \section{Exceptional DHOST~Ia subclass}

  Formulas  \eqref{nov18-20-1} and   \eqref{nov18-20-2} suggest
  that the case $f=g$ is special. Indeed, making use of
  eq.~\eqref{nov9-20-1} one finds that in this case
the matrix $G^{-1}F$ is triangular  {\it for any
    value of $Y$}, so that
  one of the sound speeds
  remains unmodified,
  $ c_{\mathcal{S} , +}^2=c_m^2=P_Y/Q$,
  while another is not necessarily
  superluminal  ($ c_{\mathcal{S} , -} = c_{\mathcal{S} , 0} + \mathcal{O}(Y)$
  for small $Y$).
  For luminal extra scalar with $c_m=1$, the condition $f=g$ gives a constraint
  $\mathcal{D}\dot{\pi} = - (3\mathcal{G_T}+\mathcal{F_T})\Delta$
    on the DHOST Lagrangian, or,
  explicitly,
  \be
A_3 = \dfrac{2(A_1-2F_{2X})(A_1 X - 2 F_2)}{X(3 A_1 X - 4 F_2)} \; .
\label{nov18-20-5}
\ee
This is the exceptional subclass of DHOST~Ia theories in which
adding extra luminal scalar field does not necessarily leads to
superluminality. Note that this subclass includes the theory
with $A_1 = 2F_{2 X}$ and $A_3=0$, which is Horndeski, and does {\it not}
include beyond Horndeski (GLPV) theories with
$XA_3 = 2A_1 - 4 F_{2 X}$ (the latter relation is inconsistent with
\eqref{nov18-20-5} for $\mathcal{G_T} \neq 0$, 
{see eq.~\eqref{eq:App_Gt}}).
This observation is in agreement with Ref.~\cite{Mironov_subl_kess}.

\section{Conclusion}
\label{sec:discussion}

We have shown that in a general DHOST~Ia theory, even tiny
amount of kinetic energy $Y>0$ of {the} rolling scalar field $\chi$ 
with the Lagrangian~\eqref{nov7-20-1} and  flat-space
sound speed 
{$c_m=1$} immediately induces 
superluminality of perturbations about the 
cosmological DHOST~{Ia} background. 
This result is valid for any stable, spatially
flat cosmological background in DHOST Ia theory with any choice
of Lagrangian functions in the action~\eqref{eq:DHOST},
except for the special subclass of theories defined by
eq.~\eqref{nov18-20-5}.

If one insists on the absence of superluminality,
one has two logical possibilities. One of them is
that in scalar-tensor theories with multiple scalars, 
none of the scalar fields has  conventional kinetic term
as long as there is at least one field 
of  DHOST Ia type. Another is to stick to the exceptional subclass
\eqref{nov18-20-5} of DHOST~Ia theories and allow for
conventional scalars. It remains to be understood, however,
whether the second class of theories is consistent
in non-cosmological backgrounds.

\section{Acknowledgements}

The work on the main part of this paper has been supported
by Russian Science Foundation grant 19-12-00393. S.M. and V.V. have been
supported by the Foundation for the Advancement of 
Theoretical Physics and Mathematics “BASIS” in their work on Appendix.

\section*{Appendix}
\label{sec:App}

In this Appendix we present the calculation of 
the quadratic action for perturbations~\eqref{eq:quadr_action_DHOSTIa+k}
for the system of DHOST Ia + k-essence described by the sum 
of actions~\eqref{eq:DHOST} and~\eqref{eq:action_setup_kess}.
Our results here generalize the derivation
given in Ref.~\cite{Mironov_subl_kess} 
for the beyond Horndeski subclass and
agree with the existing results in Refs.~\cite{DHOST_EFT,DHOST_cosmo}
wherever there is an overlap.

We work with the ADM parametrization  \eqref{nov26-20-1},
\eqref{nov26-20-2}
for perturbations about
isotropic and homogeneous background with {spatially}
flat geometry.
The background has $\dot{\pi} \neq 0$ and $\dot{\chi} \neq 0$.
We work in the unitary gauge
$\delta \pi =0$, while perturbation of k-essence
$\delta \chi$ is non-zero.

We begin with the quadratic part of k-essence action
\eqref{eq:action_setup_kess}:
\begin{multline}
\label{eq:App_quadr_action_kess}
S^{(2)}_{\chi} = \int \mathrm{d}t\,\mathrm{d}^3x \,a^3 \left[ 
Y Q \,\alpha^2 - 2 \dot{\chi} Q \, \alpha\dot{\delta\chi} 
+ 2\dot{\chi} P_Y \, \delta\chi\dfrac{\nabla^2\beta}{a^2}
+ Q\, \dot{\delta\chi}^2 - P_Y\, \dfrac{(\nabla\delta\chi)^2}{a^2} 
\right.\\ \left. 
- 6 \dot{\chi} P_Y\, \dot{\zeta}\delta\chi 
+ (P_{\chi}-2 Y P_{\chi Y})\,\alpha\delta\chi
+\Omega \, \delta\chi^2
\right],
\end{multline}
where  
$$Q = P_Y + 2 Y P_{YY,}$$
$$\Omega = P_{\chi\chi}/2 - 3 H \dot{\chi} P_{\chi Y} - Y P_{\chi\chi Y}  - \ddot{\chi} (P_{\chi Y} + 2 Y P_{\chi YY}).$$
Let us {remind}
that hereafter we do not use the background equations of motion
when deriving the terms with derivatives and also terms involving
$\alpha$ and $\beta$ only. However, we did use
background equations of motion to obtain the terms without
derivatives in~\eqref{eq:App_quadr_action_kess} and, in particular,
to see that the term 
$\zeta\cdot \delta \chi$ vanishes. There are also terms with
$\zeta^2$ and $\alpha \zeta$ in~\eqref{eq:App_quadr_action_kess} which
are not written because these terms vanish in the complete
quadratic action $S_\pi^{(2)} + S_\chi^{(2)}$ upon using the background equations
of motion. 
Anyway, we keep the non-derivative
terms for completeness only; they are
irrelevant for obtaining our main results.

Now we turn to the quadratic action for DHOST Ia sector, which in 
the unitary gauge reads:
\begin{align}
S^{(2)}_{\pi}=\int &\mathrm{d}t\,\mathrm{d}^3x \,a^3\Bigg[\left(\dfrac{\mathcal{{G}_T}}{8}\left(\dot{h}^T_{ik}\right)^2-\dfrac{\mathcal{F_T}}{8a^2}\left(\partial_i h_{kl}^T\right)^2\right)+
\left(-3\mathcal{{G}_T}\dot{\zeta}^2+\mathcal{F_T}\dfrac{(\nabla\zeta)^2}{a^2}-2(\mathcal{G_T}+\mathcal{D}\dot{\pi})\alpha\dfrac{\nabla^2\zeta}{a^2}
\right.\nonumber\\&\left.
+\Sigma\alpha^2+6\Theta\alpha\dot{\zeta}-2\Theta\alpha\dfrac{\nabla^2\beta}{a^2}
+2\mathcal{{G}_T}\dot{\zeta}\dfrac{\nabla^2\beta}{a^2}\right) +
{\left(\Xi \dfrac{(\nabla\alpha)^2}{a^2} +\Gamma \dot{\alpha} \dfrac{\nabla^2\beta}{a^2} - 3\Gamma \dot{\alpha}\dot{\zeta} - \dfrac{3\Gamma^2}{4\mathcal{G_T}} \dot{\alpha}^2 \right)}
\Bigg],
\label{eq:App_quadr_action_DHOSTIa}
\end{align}
where $(\nabla\zeta)^2 = \delta^{ij} \partial_i \zeta \partial_j \zeta$,
$\nabla^2 = \delta^{ij} \partial_i \partial_j$,
$\mathcal{G_T}$ and  $\mathcal{F_T}$ are given by
\eqref{eq:App_Gt} and \eqref{eq:App_Ft}, respectively,
and
\begin{subequations}
\label{eq:App_list_of_coefficients}
\begin{align}
& \mathcal{D} = -2 A_1 \dot {\pi} + 4 F_ {2 X} \dot {\pi},\\
& \Gamma = X (-2 A_1+4 F_{2X}+A_3 X),\\
& \Xi = \frac{\Gamma}{2 \mathcal{G_T}^2} 
\left[8 F_2^2-2 F_2 (5 A_1+6 F_{2X}) X+(A_3 F_2+16 A_1 F_{2X}) X^2\right],\\
& \Theta  =  (-A_{3X}+A_5) \ddot{\pi} \dot{\pi} X^2-\dot{\pi} (F_{2\pi}+\ddot{\pi}
(-3A_1+6 F_{2X}))-2 F_2 H
\nonumber\\
&+ X (3 A_1+2 F_{2X}) H + X^2 \left(-2 A_{1X} H + \frac{3}{2}
 (4 A_{1X}+A_3) H \right)
\nonumber\\
&-\dot{\pi} X \left( \ddot{\pi}
(- 2 A_{1X}+ \frac{3}{2} A_3- A_4+4 F_{2XX})+2 F_{2\pi X}+K_X \right) \; 
\label{eq:App_Theta_action},
\\
&\Sigma = 6 F_2 H^2-2 \ddot{\pi} (A_{5\pi} X+A_{5X X} \ddot{\pi}) \dot{\pi}^8
-\dot{\pi}^7 (2 A_{5X} \dddot{\pi}+3 A_{3\pi X} H+6 A_{5X} \ddot{\pi} H)
\nonumber\\
&+6 \dot{\pi} [ F_{2\pi} H-2 \ddot{\pi} (A_1-2 F_{2X}) H]+\dot{\pi}^6 [-2 (A_{3\pi X}+A_{4\pi X}+4 A_{5\pi}) \ddot{\pi}
  \nonumber\\
    &-(2 A_{3X X}+2 A_{4X X}+13 {A_5X}) \ddot{\pi}^2-3 A_{3X} \dot{H}
  -3 (4 A_{1X X} + 3 A_{3X}) H^2]
\nonumber\\
  &+\dot{\pi}^2 [-3 (A_3+A_4)
    \ddot{\pi}^2-12 \dot{H} (-A_1+2 F_{2X})+ F_{X} 
    -42 F_{2X} H^2-K_{\pi}]
\nonumber\\
& +\dot{\pi}^4 [-6 (A_{3\pi}+A_{4\pi})
    \ddot{\pi} 
    -(9 A_{3X}+9 A_{4X}+12 A_5) \ddot{\pi}^2-3 \dot{H} (-2 A_{1X}+3 A_3+4 F_{2X X})+
    2  F_{X X} \nonumber\\
& -36 A_{1X} H^2-27 A_3 H^2-24 F_{2X X} H^2-K_{\pi X}]
-\dot{\pi}^3 \;\dddot{\pi} [6 (A_3+A_4) \dddot{\pi}
  \nonumber\\
  &+3 \ddot{\pi} (2 A_{1X}+3 A_3+6 A_4-4 F_{2X X}) H+6 H (-2 A_{1\pi}-F_{2\pi X}-2 K_X)]
\nonumber\\
&-\dot{\pi}^5 \left[2 (A_{3X}+A_{4X}+4 A_5)
+3 (A_{3X}+2 A_{4X}+8 A_5) \ddot{\pi} H+3 H (-2 A_{1\pi X}+3 A_{3\pi}-2 K_{X X})\right]. \label{eq:App_Sigma_action}
\end{align}
%
\end{subequations}
Here the functions $A_4$ and $A_5$ are given by \eqref{a4_A} and
\eqref{a5_A}, respectively. For the same reason as before we do
not write terms with $\zeta^2$ and $\alpha\zeta$: they cancel out
in the total quadratic action upon using background equations of motion.

 We keep the notation of 
the coefficients in~\eqref{eq:App_quadr_action_DHOSTIa} similar to
that in Ref.~\cite{Mironov_subl_kess} 
to compare and contrast the 
quadratic actions for Horndeski and beyond Horndeski theories on the one hand,
and DHOST Ia theories on the other.
In particular, the terms in the last parentheses in
eq.~\eqref{eq:App_quadr_action_DHOSTIa} are 
new as compared to the beyond Horndeski case, in which
the coefficients $\Xi$ and $\Gamma$ vanish,  cf. 
Refs.~\cite{KobaReview,Mironov_subl_fluid,Mironov_subl_kess}. 

Clearly, adding the scalar field $\chi$ to DHOST theory does not
give anything qualitatively new for the tensor sector, so from now on we 
focus on the scalar sector.
As a consequence of the degeneracy feature of DHOST Ia theory, the kinetic
matrix
for the scalar DOFs in 
the action~\eqref{eq:App_quadr_action_DHOSTIa} has 
vanising determinant and, hence,
the terms $\dot{\zeta}^2$, $\dot{\alpha}\dot{\zeta}$ 
and $\dot{\alpha}^2$  combine into  perfect square. The latter property
enables one to introduce, instead of $\zeta$,
a new variable, which is the dynamical
scalar DOF in DHOST Ia:
\be
\label{eq:App_new_zeta}
\tilde{\zeta} = \zeta - \Delta \alpha,
\ee
where
\be
\label{eq:App_beta1}
\Delta = -\dfrac{\Gamma}{2\mathcal{G_T}},
\ee
and its explicit expression is given by \eqref{nov19-20-1}.
 In terms of 
the new variable $\tz$, the scalar part of the 
action~\eqref{eq:App_quadr_action_DHOSTIa} reads
\begin{multline}
\label{eq:App_quadr_action_DHOSTIa_newzeta}
\mathcal{S}^{(2)}_{\pi} = \int \mathrm{d}t\,\mathrm{d}^3x \,a^3\Bigg[
-3\mathcal{{G}_T}\dot{\tz}^2 +\mathcal{F_T}\dfrac{(\nabla\tz)^2}{a^2}
-2(\mathcal{G_T}+\mathcal{D}\dot{\pi}+
\mathcal{F_T}\Delta)\alpha\dfrac{\nabla^2\tz}{a^2} \\ 
+ \tilde{\Sigma}\alpha^2
+ 6 \tilde{\Theta} \alpha\dot{\tz}
-2 \tilde{\Theta} \alpha\dfrac{\nabla^2\beta}{a^2}
+2\mathcal{{G}_T}\dot{\tz}\dfrac{\nabla^2\beta}{a^2}
\Bigg],
\end{multline}
{where  
\begin{subequations}
\label{eq:App_theta_sigma}
\begin{align}
&\tilde{\Theta} = \Theta - \mathcal{G_T}\dot{\Delta}, \\
&\tilde{\Sigma} = \Sigma + 3\mathcal{G_T} \dot{\Delta}^2+ 6\tilde{\Theta} \dot{\Delta} 
-\dfrac{3}{a^3}\dfrac{\mathrm{d}}{\mathrm{d}t}\Big[a^3\left(\tilde{\Theta} +\mathcal{G_T} \dot{\Delta}\right)\Delta\Big],
\end{align}
\end{subequations}
and $\Theta$ and $\Sigma$ are explicitly 
given by~\eqref{eq:App_Theta_action} and~\eqref{eq:App_Sigma_action}.
The
coefficients $\tilde{\Theta}$ and $\tilde{\Sigma}$ } play similar roles as the
coefficients $\Theta$ and $\Sigma$, respectively,
in (beyond) Horndeski case.

Now we combine DHOST Ia  and  k-essence components, and use the
variable $\tz$ instead of $\zeta$ in  $S^{(2)}_{\chi}$,
eq.~\eqref{eq:App_quadr_action_kess}. 
Variables $\alpha$ and $\beta$ enter the quadratic
action without
time derivatives (the first term in the second
line in~\eqref{eq:App_quadr_action_kess}, written in terms of $\tz$,
involves $\dot{\alpha}$, but this is taken care of by integration by
parts
), so the
variation of the total quadratic action
$S^{(2)}_{\pi}+S^{(2)}_{\chi}$ 
with respect to 
$\beta$ and $\alpha$ gives two constraint equations:
\begin{subequations}
\label{eq:App_constraints}
\begin{align}
\label{eq:App_beta}
\alpha &= \dfrac{\mathcal{G_T}\dot{\tz}+ \dot{\chi}P_Y \delta \chi}{\tilde{\Theta}}, \\
 \dfrac{(\nabla^2\beta)}{a^2} &= \dfrac{1}{\tilde{\Theta} }\left(
\left[\tilde{\Sigma} +Y\; Q
 \right]\alpha 
 - \left(\mathcal{G_T} + \mathcal{D} \dot{\pi} + \mathcal{F_T}\Delta \right) 
\dfrac{(\nabla^2\tz)}{a^2} + 3 \tilde{\Theta} \dot{\tz}\right.
\nonumber\\
&\left.
 - \left(1- 3 \dfrac{P_Y}{Q}\Delta \right) \dot{\chi} Q\, \dot{\delta \chi} 
+  \frac12 (P_{\chi} - 2 Y P_{\chi Y} - 6 \dfrac{\mathrm{d}}{\mathrm{d}t}\left(
P_Y \dot{\chi} \right) \Delta )\,\delta \chi\right) \; .
\label{eq:App_alpha}
\end{align}
\end{subequations}
Plugging these solutions for $\alpha$ and $\beta$ back into the action
$S^{(2)}_{\pi}+S^{(2)}_{\chi}$, we obtain the unconstrained
quadratic action
in terms of two dynamical variables $\tz$ and 
$\delta\chi$: 
\be
\label{eq:App_quadratic_action_final}
S^{(2)}_{\pi+\chi} = \int \mathrm{d}t\,\mathrm{d}^3x \,a^3
\left[G_{AB} \dot{v}^A \dot{v}^B 
- \dfrac{1}{a^2} F_{AB} \nabla_i\,{v^A} \nabla^i\,{v^B}+ 
\Psi_1\dot{\tz}\delta\chi + \Psi_2 \delta\chi^2\right],
\ee
where notations are the same as in \eqref{eq:quadr_action_DHOSTIa+k}, i.e.,
$A,B = 1,2$ and $v^1 = \tz$, $v^2 = \delta\chi$, while matrices
$G_{AB}$ and $F_{AB}$ are given by \eqref{nov9-20-1}, with
$\mathcal{{G}_S}$,  $\mathcal{{F}_S}$ defined in
\eqref{eq:App_Gs} and \eqref{eq:App_Fs}.
Even though the coefficients $\Psi_1$ and $\Psi_2$ in~\eqref{eq:App_quadratic_action_final} are irrelevant 
for the analysis of possible ghost and gradient instabilities, as well as
(super)luminality,
we give their explicit form here for completeness:
\begin{subequations}
\begin{align}
\label{eq:App_psi}
&\Psi_1 = \dfrac{\mathcal{G_T}}{\tilde\Theta^2}\left[ 2 \dot{\chi} P_Y  
(\tilde\Sigma + Y R) + \tilde\Theta 
\left(P_{\chi} - 2 Y P_{\chi Y} +
\dfrac{6 \Delta}{a^3} \dfrac{\mathrm{d}}{\mathrm{d}t}\left[a^3 P_Y \dot\chi\right]\right)
\right] , \\
&\Psi_2 = \Omega + \dfrac{\dot\chi P_Y }{\tilde\Theta^2}\left[  \dot{\chi} P_Y  
(\tilde\Sigma + Y R) + \tilde\Theta 
\left(P_{\chi} - 2 Y P_{\chi Y} +
\dfrac{6 \Delta}{a^3} \dfrac{\mathrm{d}}{\mathrm{d}t}\left[a^3 P_Y \dot\chi\right]\right)
\right] 
\nonumber\\
&\hspace{7cm}
+ \dfrac{6 \Delta}{a^3} \dfrac{\mathrm{d}}{\mathrm{d}t}\left[ \dfrac{Y P_Y(R- 3 P_Y \Delta)}{\tilde\Theta}\right],
\end{align}
\end{subequations}
where $\Omega$ is defined in~\eqref{eq:App_quadr_action_kess}.

Let us briefly discuss
the stability conditions of the scalar sector 
in the combined system of DHOST Ia and 
k-essence, without assuming that the background value of $Y$ is small.
These conditions are somewhat 
more involved than those for the pure DHOST Ia, 
cf. eq.~~\eqref{eq:quadr_action_DHOSTIa-noscalar}, and amount
to positive definiteness of both kinetic matrices $G_{AB}$
and $F_{AB}$ in~\eqref{eq:App_quadratic_action_final}.  
Explicitly, one has
\be
\begin{aligned}
\label{eq:App_stability_scalar_2}
& \mathcal{{G}_S}>0,  \;\;\; \mathcal{{F}_S} > 0, \\
& P_Y > 0, \;\;\; Q > 0, \\
1+ \Lambda \Delta > 0, &\;\;\;
\mathcal{{F}_S} - Y P_Y\dfrac{\left(\mathcal{G_T} + \mathcal{D} \dot{\pi}+ \mathcal{F_T}\Delta\right)^2}{\tilde{\Theta}^2} > 0,
\end{aligned}
\ee
where
$$
\Lambda = \dfrac{6\mathcal{G_T}^2}{\tilde{\Theta}^2} \dfrac{Y P_Y}{\mathcal{{G}_S}} \left(1-\dfrac32\dfrac{P_Y}{Q}\Delta\right).
$$
The third line
in~\eqref{eq:App_stability_scalar_2} is characteristic of the 
combined theory, while the first two lines coincide with the 
stability conditions
for separate DHOST Ia theory and k-essence, respectively.  Note
that similar results for beyond Horndeski subclass + k-essence 
in Ref.~\cite{Mironov_subl_kess}
are recovered upon setting $\Delta = 0$ 
(see eq.~\eqref{eq:App_beta1}
and recall that $\Gamma$ vanishes in (beyond) Horndeski theories).

Finally, let us compare 
the scalar sound
speeds $c_{\mathcal{S},\, \pm}^2$ in the
{general}
DHOST Ia + k-essence system
with those
in beyond Horndeski theory~\cite{Mironov_subl_kess}, still without assuming
that $Y$ is small. Sound speeds in DHOST Ia + k-essence theory are obtained
from eq.~\eqref{eq:det}:
\be
\label{eq:App_speeds_schematic}
c_{\mathcal{S}\, \pm}^2 =  \dfrac{1}{1+\Lambda\Delta} \left(\dfrac12 (c_m^2 + \mathcal{A})
\pm  \dfrac12 \sqrt{(c_m^2 - \mathcal{A})^2 + \mathcal{B}}\right),
\ee
where
\begin{subequations}
  \label{nov29-20-10}
\begin{align}
\label{eq:App_AB}
&\mathcal{A}= \frac{\mathcal{{F}_S}}{\mathcal{{G}_S}} -
\frac{YP_Y}{\mathcal{\tilde{G}_S}} \, \frac{\mathcal{G_T}(\mathcal{G_T} + 2 \mathcal{D}\dot{\pi}- \Delta \mathcal{M})}{ \tilde{\Theta}^2} \; ,\\
&\mathcal{B}= 4c_m^2 \left(\frac{YP_Y}{\mathcal{{G}_S}}
\frac{(\mathcal{D} \dot{\pi} +\mathcal{F_T}\Delta)^2 +\Delta \mathcal{G_T}( \mathcal{M}+2\mathcal{F_T})}{ \tilde{\Theta}^2} - 
\Lambda \Delta  \left[ \dfrac{\mathcal{{F}_S}}{\mathcal{{G}_S}} - 
\dfrac{Y P_Y}{\mathcal{{G}_S}} \dfrac{\left(\mathcal{{G}_T}+\mathcal{D}\dot{\pi}+ \mathcal{F_T}\Delta\right)^2}{\tilde{\Theta}^2}\right]\right)\; ,
\label{eq:App_B}
\end{align}
\end{subequations}
and
\be
\label{eq:App_M}
 \mathcal{M} = 6(\mathcal{{G}_T}+\mathcal{D}\dot{\pi}+ \mathcal{F_T}\Delta)c_m^2 -2 \mathcal{F_T}.
\ee
As we  mentioned above, to restore the beyond Horndeski case
it is sufficient to set $\Delta=0$ in eqs.~{\eqref{nov29-20-10}}.
We see that there is a considerable 
difference between the beyond Horndeski and general DHOST Ia cases: 
in the former, the pre-factor in \eqref{eq:App_speeds_schematic}
is equal to 1, while
the coefficient $\mathcal{B}$ reads 
(cf. Ref.~\cite{Mironov_subl_kess})
\be
\label{eq:App_B_Horndeski}
\mathcal{B}= 4c_m^2 \frac{YP_Y}{\mathcal{{G}_S}}
\frac{(\mathcal{D} \dot{\pi})^2}{ {\Theta}^2},
\ee
i.e., it  is manifestly positive
for  stable and rolling background
($\mathcal{G_S}, P_Y > 0$ and $Y>0$); then  eq.~\eqref{eq:App_speeds_schematic}
immediately
gives $c_{\mathcal{S}\, +}^2 > c_m^2$, which 
implies superluminality for $c_m =1$.
For DHOST Ia the situation is less transparent: 
even if the stability conditions
are satisfied,
the coefficient $\mathcal{B}$ in eq.~\eqref{eq:App_B} is not
necessarily positive and
the pre-factor in \eqref{eq:App_speeds_schematic}
is different from 1.
So, for finite values of $Y$, the  (super)luminality issue
is not straightforwardly analyzed.


\end{document}